\documentstyle[12pt]{article}
\makeatletter
\def\section{\@startsection {section}{1}{\z@}{3.5ex plus 1ex minus 
 .2ex}{2.3ex plus .2ex}{\large\bf}}
\def\lsim{\mathrel{\mathpalette\gl@align<}}
\def\gsim{\mathrel{\mathpalette\gl@align>}}
\def\gl@align#1#2{\lower.6ex\vbox{\baselineskip\z@skip\lineskip\z@
    \ialign{$\m@th#1\hfil##\hfil$\crcr#2\crcr\sim\crcr}}}
\makeatother
\newcommand{\newsection}[1]{
\setcounter{equation}{0}
\section{#1}
}
\newcommand{\be}{\begin{equation}}
\newcommand{\ee}{\end{equation}}
\newcommand{\bea}{\begin{eqnarray}}
\newcommand{\eea}{\end{eqnarray}}

\def\ie{\hbox{\it i.e.}}

\def\xd{x_{\Delta}}
\def\xc{x_c}
\begin{document}
\begin{flushright}
OCHA-PP-90
\end{flushright}
\begin{center}
{\large{\bf Classicalization of Quantum Fluctuation \break 
            in Inflationary Universe 
       }
}
\end{center}

\begin{center}
{\sc Hiroto Kubotani
\footnote{Electronic address:kubotani@aquarius.nara-edu.ac.jp \\
\hspace*{6.0mm}Present address:Dept. of Math., 
Nara Univ. of Education, Nara 630, Japan~~~~~~~~}
}
\end{center}

\begin{center}
{\sc and}
\end{center}

\begin{center}
{\sc Tomoko Uesugi, 
\footnote{Electronic address:tomoko@phys.ocha.ac.jp}
Masahiro Morikawa, 
\footnote{Electronic address:hiro@phys.ocha.ac.jp}
Akio Sugamoto
\footnote{Electronic address:sugamoto@phys.ocha.ac.jp}
}
\end{center}

\begin{center}
{\it Department of Physics \break
Ochanomizu University,~Tokyo 112,~Japan}
\end{center}

\vfill
\begin{abstract}
We discuss the classicalization of a quantum state induced by an 
environment in the inflationary stage of the universe.
The classicalization is necessary for the homogeneous ground state to become 
classical non-homogeneous one accompanied with the statistical fluctuation, 
which is a plausible candidate for the seeds of structure formation.
Using simple models, we show that 
i) the two classicalization criteria, the classical correlation and quantum 
decoherence, are simultaneously satisfied by the environment and that 
ii) the power spectrum of the resultant statistical fluctuation depends upon 
the detail of the classicalization process.
Especially, the result ii) means that, taking account of the classicalization 
process, the inflationary scenario does not necessarily predict the unique 
spectrum which is usually believed.

\noindent
{\it keywords}:
inflation, squeezed state, spectrum of fluctuation
\end{abstract}

%
\newsection{Introduction}
%
%
%
\par
Inflationary scenario 
\cite{sato81_albrecht82}\cite{linde90} 
was introduced in order to overcome the difficulties which the standard big 
bang model holds, \ie ~the horizon problem, the flatness problem etc .
In addition to the advantage, it yields the many favorite 
features like giving an answer to the monopole problem.
Further, most attractive point for cosmologists may be the possibility 
that the exponential growth of the universe brings the 
nature of the quantum state into the macroscopic one of the present universe.
The idea that quantum fluctuations evolve into 
classical fluctuations which are expected to be the seeds of galaxies, cluster of galaxies, and large scale structures was 
proposed immediately after the proposal of inflation
\cite{hawking82_guthpi82}.

Then the quantum-to-classical transition of the whole universe at 
the inflationary stage has been further discussed by many authors.
Among previous works, there has been a confusion on the appearance of 
classical properties in quantum states, \ie ~the confusion exists between the 
classical correlation and the loss of quantum coherence
\cite{halliwell89}\cite{morikawa90}.
The former is the criterion that wave functions  
establish the classical correlation in the phase space of the system.
This property is necessary for wave functions to recover the well-defined peak 
in the phase space
\cite{hartle86}\cite{halliwell87}.
The latter is one that quantum coherence is lost and the superposition of 
states is forbidden.
Guth and Pi \cite{guthpi85}, for example,  
concluded that a squeezed state behaves classically.
Then, they identified the two-point correlation function of  
the quantum field with the statistical two-point correlation function 
averaged over the ensemble.
Indeed, they argued the classicalization based on the classical correlation.
In principle, however, 
quantum fluctuations are different from statistical fluctuations.
The spatially homogeneous quantum state itself has no intrinsic inhomogeneous 
properties.  
The standard interpretation of wave functions only tells us that, 
even if a spatially invariant observable like a power spectrum 
of a field is observed, it is reduced to one of the superposed 
homogeneous states and never be changed into a inhomogeneous one.
On the other hand, the statistical mechanics stands upon the existence 
of an ensemble.
Total system is divided into many subsystems.
In cosmology, 
initial homogeneity of the total universe is necessary to be broken. 
Therefore, 
the statistical argument cannot be allowed 
until the state of the universe becomes inhomogeneous classical one.

As long as the quantum state of the universe remains quantum, it remains 
homogeneous and has nothing to do with statistical fluctuations.
We need to consider an additional process which reduces quantum fluctuations 
to statistical fluctuations.
A standard way to describe the natural evolution from purely quantum system 
to the system with classical nature 
is to introduce the environmental degrees of freedom which couple with the 
system of interest and coarse grain the environment.  
As a result, the classical behavior is gradually 
achieved through the interaction with the environment.
This process is elegantly described by the reduced density matrix 
for the system of interest and the evolution equation for it 
\cite{caldeira83}.  
If the initial density matrix of a pure quantum state becomes, 
in the course of its evolution, diagonal with respect to some representation, 
the state has changed to the classicalized state, since
the off-diagonal elements of the density matrix represent the quantum 
coherence between quantum states.
The mixed state acquires statistical properties,
since the diagonal part of the density matrix represents the statistical 
information.

In this paper, we trace the evolution of the density matrix of 
initially quantum pure state in the era of the inflation 
under the effect of the environmental degrees of freedom.
Using this density matrix, we can discuss the classicalization of the 
quantum fluctuation of the early universe
\cite{sakagami88} \cite{brandenberger90} \cite{matacz93}.
For this aim, we use a massless scalar field as a simple model.
For each mode, we write down the equation of motion for the density 
matrix as the Fokker-Planck  equation form and solve it.
In the same model, Matacz \cite{matacz94} discussed the decoherence of 
the quantum fluctuation making the density matrix diagonal 
in a non-dynamical {\it ad hoc} way.
We would like to take account of the dynamics on the evolutionary process of 
the density matrix.
We will see how the fluctuation associated with the decoherence 
depends on the properties of the environment and compare the power spectrum of the resultant statistical fluctuation obtained after classicalization 
with the previously proposed results.

This paper is organized as follows.
In section 2, we discuss the quantum mechanics of an open system and 
the two independent criteria of classicalization for 
the squeezed state.
Then,  we interpret the state as a single mode state of the scalar field coupled  minimally with the gravitation.
In section 3, the evolution of the density matrix of the model under the 
background of the expanding universe is solved analytically.
Next, in section 4, we discuss the fluctuations at the final stage of the 
inflation.
Finally, we summarize our results.

%
\newsection{Quantum Open System}
%
As a typical model, we consider a harmonic oscillator coupled with 
an environment.
We are interested in the evolution of the harmonic oscillator whose 
dynamics is affected by environmental degrees of freedom.  
In order to describe a state, in this situation, it is useful to use 
the reduced density matrix $\hat{\rho}(t)$
which is obtained 
after coarse-graining the environmental degrees of freedom.
We can easily derive the equation of motion for it according to 
Schwinger\cite{schwinger61}.
When the harmonic oscillator $x$ couples with the environment, 
the equation of motion for $\rho(t)$ becomes
\bea
\dot\rho(x,x';t)&=&
\Big[  
{i \over 2m}(\partial^2_x-\partial^2_{x'})-i{m \over 2}\omega^2(x^2-x'^2)
-i(v(x)-v(x')) 
~~~~~~~~~~~~~~~~\nonumber \\
&&~~~~~~~~~~~~~~-\epsilon\cdot (x-x')(\partial_x-\partial_{x'}) 
              -\Lambda\cdot (x-x')^2
\Big]
\rho(x,x';t), 
\label{fpenv}
\eea
where 
$\rho(x,x';t)\equiv \left<x | \hat{\rho}(t) |x'  \right>,$ 
and $m$ and $\omega$ are the mass and frequency of the oscillator, 
respectively.
In this equation, there appear three fundamental effects 
from the environment:

a) dissipation effect which is represented by the friction 
coefficient $\epsilon$, 

b) fluctuation effect which is represented by the diffusion coefficient $\Lambda$, 

c) renormalization effects which force the parameters in the original Hamiltonian shift as 
$v$.  

\noindent
The former two effects, a) and b), are nonlocal and complicated 
in general.
In order to make the research of the solution 
simple, we have taken a local approximation which corresponds to 
the Markovian limit in the dissipative corrections.

In the next section, 
when we actually calculate the evolution of density matrix, 
we will consider Gaussian states.
The general formula for it is given as 
\be
\rho(x,x';t)
={\rm exp}(-\alpha^2 x_c^2+i\beta x_c x_{\Delta}-\gamma^2 x_{\Delta}^2
           +\mu x_c+i\nu x_{\Delta}+\lambda),
\label{rho1}
\ee
where $2\xc=x+x'$ and $\xd=x-x'$ and 
all the coefficients $\alpha$, $\beta, \ldots \lambda$ 
are real functions of time for the Hermiticity of $\rho(t)$.
Normalizability of $\rho(t)$ requires the functions 
$\alpha$ and $\gamma$ to be positive.
In order to obtain the evolution of the density matrix, 
we only have to integrate the equation of motion for 
$\alpha$, $\beta, \ldots \lambda$.
However, the criteria and the measure of the classicality of this quantum 
state is not straightforward and further discussion is necessary.  
Many criteria for the classicalization
have been proposed so far
\cite{machida80_zurek81}.  
In the remaining of this section, 
we shall summarize the proposals from the point of 
the two conditions of classicality: classical correlation and 
quantum decoherence, and discuss availability of them
\cite{morikawa90}.

One possible measures of classicality is how well the classical peak
in the phase space is recovered.  
Quantum states cannot have finer resolution in the phase space than the Planck constant $\hbar$. 
Accordingly, the sharpness of correlation of the coordinate to 
the corresponding momentum can be a classicality measure. 
The Gaussian state, Eq.\ref{rho1}, is transformed into 
the Wigner representation as 
\be
W(x_c,p;t)={\sqrt{\pi} \over \gamma}{\rm exp}
[-{(\beta x_c+\nu-p)^2 \over 4 \gamma^2}-\alpha^2 x_c^2+\mu x_c+\lambda]. 
\label{wigner}
\ee
In this state, the classical correlation will be strongly established 
if the system is squeezed.
The classicality measure may be given by the relative sharpness of 
the peak in the phase space, 
\be
\delta_{CC}=2 \alpha \gamma /|\beta|.
\label{delcc}
\ee
The condition $\delta_{CC} \ll 1$ means strong classical correlation.
This measure is dependent on the representation (choice of canonical variables).

One of the direct effects of the environment which couples to a system 
is the random perturbations on the system.
It makes the quantum coherence of the system disappear.
This decoherence may be an another criterion for the classicality.
As an actual measure for the density matrix 
$\hat{\rho}$ , we introduce the linear entropy 
$\delta_{QD}\equiv{\bf Tr}\hat{\rho}^2$ or the 
statistical entropy $S\equiv-{\bf Tr}\hat{\rho} {\bf ln}\hat{\rho}$.
The quantity $\delta_{QD}$ becomes 1 for a pure state and reduces toward 0 
as the quantum coherence is lost.  
On the other hand, the entropy $S$ becomes 0 for a pure state and increases to 
infinity if the quantum coherence is destroyed.  
These are independent of the representation.   
For the Gaussian state Eq.\ref{rho1} with  $\mu=0$, we have
\bea
\delta_{QD}&=&{1 \over 2}{\alpha \over \gamma}, 
\label{delqd} \\
S&=&{\rm ln}(({\alpha \over \gamma})^{-1}+{1 \over 2})
    -(({\alpha \over \gamma})^{-1}-{1 \over 2})
   {\rm ln}(1-{1 \over 2}{\alpha \over \gamma})
          /(1+{1 \over 2}{\alpha \over \gamma}).
\label{entropy}
\eea
The increase of $S$ indicates the decrease of $\delta_{QD}$, 
since $d \delta_{QD} /d S$ is always negative.  
Therefore $\delta_{QD}$ and $S$ are consistent as a measure of decoherence.
Before closing this section, we discuss the relation of the environmental 
effect to this criterion.
From the evolution equation for the density matrix, we can 
derive the evolution equation for the decoherence measure $\delta_{QD}$ as
\be
\dot \delta_{QD}
=\epsilon \delta_{QD} -2 \Lambda {\bf Tr}(\hat{x}^2 \hat{\rho}^2
-(\hat{x}\hat{\rho})^2).  
\label{eomqd}
\ee
It shows that the dissipation term (proportional to $\epsilon$ ) promotes 
quantum coherence, whereas  the fluctuation term (proportional to $\Lambda$) 
reduces the coherence.  
It is thus clear that the balance of dissipation and fluctuation determines 
the final equilibrium value of $\delta_{QD}$ if it exists.

%
\newsection{Quantum State in the Expanding Universe}
%
In this section, we discuss the evolution of a quantum state in the inflationary universe using the equation of motion 
for an open system developed in the previous section.
The expansion of the universe makes 
a quantum state unstable.
It makes the initial ground state squeeze and creates a classical correlation.
The squeezing of a state itself, however, does not yield statistical 
fluctuations.
From the point of generation of classical fluctuations, 
the decay of quantum coherence is indispensable.

In order to proceed our discussion concretely, we consider a scalar 
field coupled with  environmental degrees of freedom.
Lagrangian of the free part of the scalar field is given by
\be
{\cal L}=\int dx^4 {1 \over 2}[-g(x)]^{1 \over 2}
\{g^{\mu \nu}(x)\phi(x)_{,\mu}\phi(x)_{,\nu}-[m^2+\xi R(x)]\phi(x)^2\}, 
\label{action}
\ee
where $\xi $ is a coupling constant of the field to the scalar curvature $R$.
Suppose that the spatially homogeneous part of this field yields inflationary background 
\be
ds^2= a^2(d\eta^2-d{\bf x} \cdot d{\bf x}) ,
\label{metric}
\ee
where $a=-1/H\eta~(-\infty < \eta <0)$.
Then, the Lagrangian density of the field is reduced to 
\be
L(x)={1 \over 2}a^2[{\dot{\phi}}^2-\nabla \phi \cdot \nabla \phi 
      -(m^2 a^2+6 \xi {\ddot{a} \over a})\phi^2],
\label{homlag}
\ee
where over dot denotes a derivative with respect to conformal time $\eta$.
After the conformal transformation of the scalar field, 
\be
\phi \longrightarrow \psi = a \phi ,
\label{conftran}
\ee
the Lagrangian density becomes 
\be
L(x)={1 \over 2}[{\dot{\psi}}^2-\nabla \psi \cdot \nabla \psi 
      -(m^2 a^2+(6 \xi -1){\ddot{a} \over a})\psi^2  
      -{d \over d\eta}({\dot {a} \over a}\psi^2)],
\label{homlag2}
\ee
where we will drop the last term: the exact differential term.
We expand the scalar field $\psi $ in the form of Fourier series in 
the comoving spatial volume $L^3$, 
\be
\psi(x)=\sqrt{{2 \over L^3}} 
\sum_{{\bf k}}
[q_{\bf k}^e {\rm cos}({\bf k} \cdot {\bf x}) 
                        +q_{\bf k}^o {\rm sin}({\bf k} \cdot {\bf x})],
\label{fexpand}
\ee
where $q_{\bf k}^e$ and $q_{\bf k}^o$ are, respectively, 
the cosinoidal and sinoidal Fourier 
coefficients with spatial wave number ${\bf k}$.
We notice that 
we use real Fourier series to keep the field variables real.
Now, the Lagrangian becomes 
\be
{\cal L} =\int d\eta {1 \over 2} 
\sum_{\alpha =\{e,o\}}\sum_{\bf{k}}
   [(\dot{q_{\bf{k}}^\alpha })^2
  -(k^2+m^2 a^2+(6\xi-1){\ddot{a} \over a})(q_{\bf{k}}^\alpha)^2 .
\label{klag}
\ee
It shows that the system is reduced to an infinite 
ensemble of harmonic oscillators 
with time-dependent potential through the scale factor $a$.

Next we will introduce the coupling of the environment 
to the $\psi$ field phenomenologically.
Another quantum field is explicitly introduced as the environment  
in the model by Sakagami\cite{sakagami88}. 
Our model corresponds to the Gaussian approximation to the effective 
action and Markovian approximation to the nonlocal effect of the 
environment.
Therefore, according to section2, 
we only have to consider the equation of motion 
\bea
\dot\rho(q_{\bf k}^\alpha,\acute{q}_{\bf k}^\alpha;\eta)&=
\Big[  i{\mathstrut 1 \over \mathstrut 2}(\partial^2_q-\partial^2_{\acute{q}})
&-i{\omega^2 \over 2}(q_{\bf k}^{\alpha ~2}-\acute{q}_{\bf k}^{\alpha ~2})
\nonumber \\
&&-\epsilon (\eta) q_\Delta (\partial_q-\partial_{\acute{q}}) 
-\Lambda (\eta) q_\Delta^2
\Big]
\rho(q_{\bf k}^\alpha,\acute{q}_{\bf k}^\alpha;t), 
\label{fpinf}
\eea
where 
\be
\omega^2(\eta) \equiv k^2+m^2/H^2 \eta^2+2(6\xi-1)/\eta^2 
               ~~~~~~~(k^2=\bf{k} \cdot \bf{k}) 
\label{omega2}
\ee
and $\partial_q$ denotes the partial derivative with respect to 
the filed variable $q_{\bf k}^\alpha$ and $q_\Delta =q_{\bf k}^\alpha-
\acute{q}_{\bf k}^\alpha$.
Here, $\epsilon(\eta)$ and $\Lambda(\eta)$ are the 
phenomenologically introduced 
dissipation and diffusion coefficients.
In the field theory, these environmental effects 
are physically originated from the back reaction of the 
particle creation of the environmental field or 
the radiative correction due to the self-interaction of the 
system itself\cite{morikawa86a}.

For simplicity, we consider a massless minimally coupled scalar field
$(m=\xi =0)$.
In this case, $\omega^2(\eta)=k^2-2/\eta^2$.
Note that the system becomes unstable when the physical wave-length $a/k$ of 
the mode exceeds the horizon size of the universe ($H^{-1}$).
As the environment, we consider the following two cases.
One is the case in which the environmental effect is only characterized by the 
phenomenological coefficients which are constant with respect to $\eta$
(model (a)).
In the other case, the scalar field  is coupled with the environment 
which is characterized by the physically constant coefficients (model (b)).
From the dimensional argument of 
$[\epsilon]=[\eta]^{-1}$ and $[\Lambda]=[\eta]^{-2}$, we have 
\bea
\rm{model~(a):} ~~~~~\epsilon(\eta)&=~~~~~~\epsilon_0,~~~~
                \Lambda(\eta) &=~~~~~~\Lambda_0    \label{modela} \nonumber\\
\rm{model~(b):} ~~~~~\epsilon(\eta)&=a (\eta)\epsilon_0,~~~~
                \Lambda(\eta) &=a^2(\eta)\Lambda_0, \label{modelb}\nonumber 
\eea
where $\epsilon_0$ and $\Lambda_0$ are constant.
We can distinguish several stages of the classicalization process of quantum 
fluctuation in the early universe.
For example, the coupling effects of the environment occurs 
1)   during the inflation, 
2)   at the reheating phase, and/or
3)   far later after the fluctuation scale reenters the horizon. 
Our model corresponds to the case 1).

We set the initial condition as the Gaussian ground state for each wave number ${\bf k}$ at $\eta \rightarrow -\infty$.
The reduced density matrix is always expressed as 
\be
\rho(q_{\bf k}^\alpha,\acute{q}_{\bf k}^\alpha;\eta)
={\rm exp}[-\alpha^2(\eta)q_c^2+i \beta(\eta) q_c q_{\Delta}
                      -\gamma^2(\eta)q_{\Delta}^2 +\lambda(\eta)],
\label{infrho}
\ee
where $q_c\equiv{1 \over 2}(q_{\bf k}^\alpha +\acute{q}_{\bf k}^\alpha)$, 
      $q_{\Delta} \equiv q_{\bf k}^\alpha-\acute{q}_{\bf k}^\alpha$.
Hereafter, we ignore the index ${\bf k}$ and $\alpha$ for simplicity.   
Fourier transforming $\rho$ from the variable $q_c$ to $r$, we have  
\be
P(q_\Delta,r;\eta)={\rm exp}(-A(\eta)q_\Delta^2-2B(\eta)q_\Delta r 
-C(\eta)r^2-D(\eta)), 
\label{denmatrix}
\ee
where 
\be
A=\gamma^2+{\beta^2 \over 4\alpha^2},~~B={\beta \over 4\alpha^2},~~
C={1 \over 4\alpha^2},~~D=-\lambda-{\rm ln}{\sqrt{\pi} \over \alpha}.
\label{abcddef}
\ee
The evolution equation for $P$ becomes, from Eq.\ref{fpinf}, 
the following simple form:
\be
{d \over d\eta}
\left( \begin{array}{c} A \\B \\ C/2 \end{array}\right)
=
\left[ \sum_{a=1}^3 f_{a} \hat{J}_{a}+f_{3} \hat{I} \right]
\left( \begin{array}{c} A \\B \\ C/2 \end{array}\right)
+
\left( \begin{array}{c} \Lambda \\ 0 \\ 0 \end{array}\right),
~~~~{d\over d\eta}D =0, 
\label{eom3}
\ee
where
\be
f_{1}=-\sqrt{2}\omega^2(\eta)+{1 \over \sqrt{2}}
~,~~~~~
f_{2}=(-\sqrt{2}\omega^2(\eta)-{1 \over \sqrt{2}})i
~,~~~~~
f_{3}=-2 \epsilon (\eta)
\label{fundef}
\ee
\be
\hat{J}_{1}=
{1 \over \sqrt{2}}
\left( \begin{array}{ccc} 0&1&0 \\ 1&0&1 \\ 0&1&0 \end{array}\right), 
\hat{J}_{2}=
{1 \over \sqrt{2}}
\left( \begin{array}{ccc} 0&-i&0 \\ i&0&-i \\ 0&i&0 \end{array}\right), 
\hat{J}_{3}=
\left( \begin{array}{ccc} 1&0&0 \\ 0&0&0 \\ 0&0&-1 \end{array}\right)
\label{matdef}
\ee
and $I$ is the $3 \times 3$ identity matrix.

Before integrating the inhomogeneous equation, Eq.\ref{eom3},
we first solve the homogeneous one (Eq.\ref{eom3} with $\Lambda =0$).
If we integrate it partially and introduce the new variable
\be
|\chi>
\equiv 
\left( \begin{array}{c} \tilde{A} \\ \tilde{B} \\ \tilde{C}/2 \end{array} 
                                                           \right)
=U(\eta)^{-1} \left( \begin{array}{c} A \\ B \\ C/2 \end{array} \right),
\label{chi}
\ee
the equation of motion is reduced to 
\be
{d \over d\eta}|\chi >=\hat{H}_{3 \times 3}|\chi>,
\label{algeeq}
\ee
where $U(\eta)={\rm exp}(\int^\eta f_3(\eta') d\eta')$ and 
$\hat{H}_{3 \times 3}\equiv \left[\sum_{a=1}^3 f_a \hat{J}_a \right] $. 
We can interpret this equation as a Schr\"{o}diger equation for a spin-1 state $|\chi>$.
The original dynamics of $\rho(t)$ is technically equivalent to the rotation of a spin.   
We can represent this state as a direct product of two independent 
spin-${1 \over 2}$ states (dotted and un-dotted spinors)
\be
|\chi>=
\left( \begin{array}{c} u_1 v_1 \\
                       (u_1 v_2 + u_2 v_1)/\sqrt{2} \\
                        u_2 v_2          \end{array} \right),
\label{relation}
\ee
where 
\be
{\bf u}=\left( \begin{array}{c} u_1 \\ u_2 \end{array} \right)
~~~~~{\rm and}~~~~~
{\bf v}=\left( \begin{array}{c} v_1 \\ v_2 \end{array} \right)
\label{spinor}
\ee
are the dotted and un-dotted spinor, respectively.
For these spinors, the 3-dimensional rotation is expressed as 
\be
{d \over d\eta} {\bf u}=\hat{H}_{2 \times 2} {\bf u}
~~~~~{\rm and}~~~~~
{d \over d\eta} {\bf v}=\hat{H}_{2 \times 2} {\bf v}
\label{spineom} 
\ee
where
\be
\hat{H}_{2 \times 2}=\sum_{a=1}^3 f_a {\sigma^a \over 2}
={1 \over 2}\left( \begin{array}{rr} f_3      & f_{-}\\
                                     f_{+}    & -f_3 
                   \end{array} \right), 
\label{su2exress}
\ee
$f_{+}=f_{1}+if_{2}$, $f_{-}=f_{1}-if_{2}$, and $\sigma^a$ is the 
Pauli matrix.
By eliminating $u\equiv u_1$({\rm or~~}$v_1$), we obtain a differential 
equation for $d\equiv u_2$({\rm or~~}$v_2$):
\be
\ddot{d}=
 [-\omega^2(\eta)+\epsilon^2+\dot{\epsilon}]d,
\label{deom}
\ee
where the over dot denotes the differentiation with respect to $\eta$.
For both cases of model (a) and (b), 
Eq.\ref{deom} reduces to the Bessel differential equation.   
General solution for $d(\eta)$ is 
\be
d(\eta)=C_1 F_1(\eta)+C_2 F_2(\eta),
\label{generald}
\ee
where 
\be
F_1(\eta)=(-\tilde{k} \eta )^{1 \over 2} J_{\nu}(-\tilde{k} \eta)
~~~~{\rm and}~~~~~
F_2(\eta)=(-\tilde{k} \eta )^{1 \over 2} N_{\nu}(-\tilde{k} \eta),
\label{sola}
\ee
and $C_1$ and $C_2$ are integration constants.
Here
\bea
\rm{model~(a):} \nu^2=&9/4~~~~~~~~~~~~~~~~~~~~~~~~~~,&
                \tilde{k}^2=k^2-\epsilon^2  \nonumber\\
\rm{model~(b):} \nu^2=&9/4+(\epsilon_0/H)+(\epsilon_0/H)^2,&
                \tilde{k}=k .\nonumber
\eea
The general solution $u \equiv u_1({\rm or~~}v_1)$ 
which corresponds to Eq.\ref{generald} is 
\be
u(\eta)=C_1 G_1(\eta)+C_2 G_2(\eta),
\label{generalu}
\ee
where 
\be
G_1(\eta)=\sqrt{2}\{\dot{F}_1-\epsilon F_1 \}
~~~~~{\rm and}~~~~~
G_2(\eta)=\sqrt{2}\{\dot{F}_2-\epsilon F_2 \} .
\label{solG}
\ee
Here we have got three independent solutions of the homogeneous equation
(Eq.\ref{eom3} with $\Lambda=0$) as 
\be
|\chi_1 >=
\left( \begin{array}{c} G_1 ^2 \\ \sqrt{2} G_1 F_1 \\ F_1^2 \end{array} \right),|\chi_2 >=
\left( \begin{array}{c} G_1 G_2 \\ (G_1 F_2 + G_2 F_1 )/\sqrt{2} \\ F_1 F_2 
                                         \end{array} \right),
|\chi_3 >=
\left( \begin{array}{c} G_2^2 \\ \sqrt{2} F_2 G_2 \\ F_2^2 \end{array} \right).
\label{sols}
\ee

Using these results, we proceed to the inhomogeneous 
equation (Eq.\ref{eom3} with $\Lambda\neq0$).
We seek the solution in the form  
\be
\left( \begin{array}{c} A \\ B \\ C/2 \end{array} \right)=
U(\eta)\sum_{i=1}^3 D_i (\eta) |\chi_i >,
\label{mitei}
\ee
where $D_i(i=1,2,3)$ are time dependent variables.
Eq.\ref{eom3} yields equations for $D_i$(i=1,2,3);
\be
{d \over d\eta }
\left( \begin{array}{c} D_1 \\ D_2 \\ D_3 \end{array} \right)
=
W^{-2} 
\Lambda (\eta)
\left( \begin{array}{c} F_2^2 \\ -2 F_1 F_2 \\ F_1^2 \end{array} \right)
U(\eta)^{-1} ,
\label{diffds}
\ee
where $W \equiv G_1 F_2-F_1 G_2=2\sqrt{2}\tilde{k}/\pi$.
The solution is 
\be
\left( \begin{array}{c} D_1 \\ D_2 \\ D_3 \end{array} \right)
=
\left( \begin{array}{c} D_{10} \\ D_{20} \\ D_{30} \end{array} \right)
+
\left( \begin{array}{c} D_{11} (\eta) \\ D_{21} (\eta) \\ D_{31} (\eta)
 \end{array} \right),
\label{eqds}
\ee
where 
\be
\left( \begin{array}{c} D_{11} (\eta) \\ D_{21} (\eta) \\ D_{31} (\eta) 
 \end{array} \right)=
{1 \over W^2}
\int_{\eta_0}^\eta d\eta \Lambda(\eta)
\left( \begin{array}{c} F_2^2 \\ -2 F_1 F_2 \\ F_1^2 \end{array} \right)
{\rm exp}(2 \int^\eta_{\eta_0} \epsilon (\eta ')d\eta ' ) .
\label{intds}
\ee
Here $D_{10}$, $D_{20}$ and $D_{30}$ are constants 
and $\eta_0$ is the time when the environmental effects set in.
We set the initial state as the ground state of the harmonic oscillator.
Therefore
\be
\left( \begin{array}{c} D_{10} \\ D_{20} \\ D_{30} \end{array} \right)
=
{\pi \over 16k}
\left( \begin{array}{c} 1 \\ 0 \\ 1\end{array} \right) .
\label{init}
\ee
As a result, substitution of Eqs.\ref{eqds} -\ref{init} into Eq.\ref{mitei} 
gives full analytic solution of Eq.\ref{eom3}.

%
\newsection{Classicalization of Fluctuations}
%
In the previous section, we have obtained the analytic expression for the 
quantum state evolution.
Using this expression, in this section, we discuss the classicalization of 
fluctuations.

First, we estimate the asymptotic formula of the density matrix 
after the wave-length exceeds the horizon size ($|\tilde{k} \eta |\ll 1$).
From Eq\ref{intds}, for the model (a), $D_{i1} (i=1,2,3)$ can be calculated as
\be
\left( \begin{array}{c} D_{11} \\ D_{21} \\ D_{31} \end{array} \right)=
{2\Lambda_0 \over \pi W^2 \tilde{k}}
\left( \begin{array}{c} 
   P(z_0)-P(z)-2\tilde{\epsilon_0} I_p   \\
   Q(z_0)-Q(z)-4\tilde{\epsilon_0} I_q   \\
   R(z_0)-R(z)-2\tilde{\epsilon_0} I_r 
   \end{array} \right) e^{-2\epsilon_0 \eta_0},
\label{anaint}
\ee
where
\bea
P(x)  &=&
[-{1 \over x}{\rm cos}^2 x
-(2\tilde{\epsilon_0}{\rm sin}^2 x +{\rm sin}(2x)+{1 \over \tilde{\epsilon_0}})
/4(\tilde{\epsilon_0}^2+1)] e^{-2\tilde{\epsilon_0}x}\nonumber \\
Q(x)  &=&
[-{2 \over x}{\rm sin} x {\rm cos} x
+(\tilde{\epsilon_0}{\rm sin}(2x)+{\rm cos}(2x))
/2(\tilde{\epsilon_0}^2+1)]e^{-2\tilde{\epsilon_0}x} \\
R(x)  &=&
[-{1 \over x}{\rm sin}^2 x
-(2\tilde{\epsilon_0} {\rm cos}^2 x-{\rm sin}(2x)+{1 \over \tilde{\epsilon_0}})
/4(\tilde{\epsilon_0}^2+1)]e^{-2\tilde{\epsilon_0}x} \nonumber
\label{anaintb}
\eea
and
\bea
I_p &=& \int^{z_0}_{z}{{\rm cos}^2 x \over x}e^{-2\tilde{\epsilon_0}x}dx 
                                                               \nonumber \\
I_q &=& \int^{z_0}_{z}{{\rm sin}x{\rm cos}x \over x}e^{-2\tilde{\epsilon_0}x}dx
                                                                         \\
I_r &=& \int^{z_0}_{z}{{\rm sin}^2 x \over x}e^{-2\tilde{\epsilon_0}x}dx .
                                                               \nonumber
\label{anaintc}
\eea
Here $\tilde{\epsilon_0}\equiv\epsilon_0/\tilde{k}$, $z\equiv |\tilde{k}\eta|$,  
and $z_0\equiv |\tilde{k}\eta_0|$.
Unfortunately $I_p$, $I_q$ and $I_r$ cannot be expressed by elementary 
functions.
We consider the weak dissipation region 
where $\epsilon_0 <k$ and, therefore,  $\tilde{k}$ is real.
Taking into account $z \ll 1 \ll z_0$, we can approximate these expressions as 
\bea
P(z_0)-P(z) &\sim& \left\{ 
\begin{array}{ll}
1/z + z_0/2  & (\tilde{\epsilon_0} z_0 \ll 1) \\
1/z + 1/4\tilde{\epsilon_0}(\tilde{\epsilon_0}^2 +1)
                                    & (\tilde{\epsilon_0} z_0  \gg 1) 
\end{array} 
        \right.  \nonumber \\
Q(z_0)-Q(z) &\sim& \left\{ 
\begin{array}{ll}
              2-{\rm sin}^2(z_0)    & (\tilde{\epsilon_0} z_0 \ll 1) \\ 
              2-1 /2(\tilde{\epsilon_0}^2+1)
                                    & (\tilde{\epsilon_0} z_0 \gg 1) 
\end {array}
        \right. \\
R(z_0)-R(z) &\sim& \left\{ 
\begin{array}{ll}
                z_0/2        & (\tilde{\epsilon_0} z_0 \ll 1) \\
               \displaystyle
               {1 \over 4\tilde{\epsilon_0}}(2-{1 \over \tilde{\epsilon_0}^2+1})                                                  & (\tilde{\epsilon_0} z_0 \gg 1)
\end{array}
        \right.  \nonumber
\label{int_a}
\eea
and
\bea
I_p &=& \left\{ \begin{array}{ll}
              \displaystyle{1 \over 2}{\rm ln}({z_0 \over z^2})
                   +O(1)
                                    & (\tilde{\epsilon_0} z_0 \ll 1) \\
                                   ~&~                               \\
              \displaystyle
               {e^{-\pi \tilde{\epsilon_0}}\over 4}
              {\rm ln}(1+{1 \over \tilde{\epsilon_0}^2})-{\rm ln}(z)
               +O(1)~~~~
                                    & (\tilde{\epsilon_0} z_0  \gg 1) \end{array} 
        \right.  \nonumber \\
I_q &=& \left\{ \begin{array}{ll}
              \displaystyle
              {1 \over 2}[{\pi \over 2}-2z]+\tilde{\epsilon_0}{\rm sin}^2 z_0
               +O(z^2,\tilde{\epsilon_0}^2 z_0) & (\tilde{\epsilon_0} z_0 \ll 1) \\ 
                                   ~&~                               \\
              \displaystyle
              {1 \over 2}{\rm arctan}({1 \over \tilde{\epsilon_0}})
               -z+O(z^3,\tilde{\epsilon_0} z^2)
                                    & (\tilde{\epsilon_0} z_0 \gg 1) \end {array}
        \right. \\
I_r &=& \left\{ \begin{array}{ll}
               \displaystyle
                {1 \over 2}{\rm ln} z_0 +O(1)
                                    & (\tilde{\epsilon_0} z_0 \ll 1) \\
                                   ~&~                               \\
               \displaystyle
                {1 \over 4}{\rm ln}(1+{1 \over \tilde{\epsilon_0}^2})
                 +O(z^2)~~~~~~~~~~~~~~~~~~
                                    & (\tilde{\epsilon_0} z_0 \gg 1)
                \end{array}
        \right. . \nonumber
\label{int_b}
\eea
For the case $\tilde{\epsilon_0}z_0 \gg 1$ where the dissipation works for 
sufficiently long time, the endpoint of the integral interval $z_0$ does not 
appear in the estimation, 
Eqs.4.4 and 4.5.
Note that, for $x \ll 1$, 
\be
|\chi_1 >\sim
\left( \begin{array}{l} 
           o(x^{-1+2\nu}) \\ o(x^{2\nu}) \\ o(x^{1+2\nu}) 
        \end{array} \right),
|\chi_2>\sim
\left( \begin{array}{l} 
           o(x^{-1}) \\ o(x^0) \\ o(x)
       \end{array} \right),
|\chi_3>\sim
\left( \begin{array}{l}
             o(x^{-1-2\nu}) \\ o(x^{-2\nu}) \\ o(x^{1-2\nu})
         \end{array} \right) .
\label{solapp}
\ee
where $\nu$ is ${3 \over 2}$.
Therefore, under the long-term dissipative case 
($|\epsilon_0 \eta_0| \gg 1$), 
we can estimate the inhomogeneous solution as 
\be
U(\eta)\sum_{i=1}^3 D_{i1} (\eta) |\chi_i >
\sim 
e^{-2 \epsilon_0(\eta-\eta_0)}D_{31}(\eta)|\chi_3>
\sim
{\pi \over 8 \tilde{k}^2}{\Lambda_0 \over \epsilon_0}|\chi_3>
\label{estimate}
\ee
We also consider the model (b).  
We can proceed the estimate in the same way as in the model (a).
Eq.\ref{intds} is written as 
\be
\left( \begin{array}{c} D_{11} \\ D_{21} \\ D_{31} \end{array} \right)
={\tilde{k} \over W^2}{\Lambda_0 \over H^2}
\left( \begin{array}{c} P'(z) \\ 
                        Q'(z) \\ 
                        R'(z) \end{array} \right)
(-\eta_0 \tilde{k})^{(2 \epsilon_0 /H)},
\ee
where 
\be
\left( \begin{array}{c} P'(z) \\ 
                        Q'(z) \\ 
                        R'(z) \end{array} \right)
=
\int_z^{z_0} {1 \over x^{1+2 \epsilon_0/H}}
\left( \begin{array}{c} N_\nu^2(x) \\ 
                        -2J_\nu(x) N_\nu(x) \\ 
                        J_\nu(x)^2 \end{array} \right)
dx.
\ee
We note that  $P'$,$Q'$ and $R'$ are still finite, 
even if $z_0$ approaches to 
infinity, since 
\bea
J_\nu(x) &\sim &\sqrt{{2\over \pi x}} {\rm cos}(x-{(2 \nu+1)\pi \over 4}),
\nonumber \\
N_\nu(x) &\sim &\sqrt{{2\over \pi x}} {\rm sin}(x-{(2 \nu+1)\pi \over 4}).
\nonumber 
\eea
Therefore, we only have to concentrate on the integration at the region 
$z \ll 1$.
In fact, we can show that, for $x \ll 1$, 
$P'(x) \sim O(x^{-2(\nu +\epsilon_0/H)})$, 
$Q'(x) \sim O(x^{-2(\epsilon_0/H)})$ and 
$R'(x) \sim R'(0)+O(x^{2(\nu-\epsilon_0/H)})$.
Here $R'(0)$ is given by the following expression: 
\be
R'(0)={1 \over 2\sqrt{\pi}}
      {\Gamma(\nu-\epsilon_0/H)\Gamma(1/2+\epsilon_0/H)
       \over \Gamma(\nu+1+\epsilon_0/H)\Gamma(1+\epsilon_0/H)},
\ee
which is of the order of $O(1)$ for $\epsilon_0/H \lsim 1$ and 
decays exponentially as $\epsilon_0/H$ increases.
Considering Eq.\ref{solapp}, we can estimate the inhomogeneous solution as 
follows:
\bea
({\eta_0 \over \eta})^{-(2\epsilon_0 /H)}\sum_{i=1}^3 D_{i1} (\eta) |\chi_i >
&\sim &
{\pi^2 \over 8 \tilde{k}}{\Lambda \over H^2}
(-\eta \tilde{k})^{(2 \epsilon_0 /H)}
(R'(0)|\chi_3>
+\left( \begin{array}{c} O(z^{-1-2\epsilon_0/H}) \\ O(z^{-2\epsilon_0/H}) \\ 
                         O(z^{1-2\epsilon_0/H}) \end{array} \right)
) \nonumber \\
&\sim &
{\pi^2 R'(0) \over 8  \tilde{k}}{\Lambda \over H^2}
(-\eta \tilde{k})^{(2 \epsilon_0 /H)}|\chi_3> .
\label{estimateb}
\eea
Summarizing the results, we have arrived at the asymptotic 
expression for the parameters of the density matrix 
in the region $|\tilde{k} \eta | \ll 1$; 
\bea
\left( \begin{array}{c} A \\ 
                        B \\ 
                        C/2 \end{array} \right)
&\sim&
{\pi \over 16 k}U(\eta)(|\chi_1>+|\chi_3>)
+{\pi \over 16 k}T(k)|\chi_3>
\nonumber \\ 
&\sim&
{\pi\over 16 k}(U(\eta)+T(k))|\chi_3>,
\label{finalsol}
\eea
where 
\be
U(\eta) = \left\{ \begin{array}{ll} 
{\rm exp}[-2\epsilon_0 (\eta-\eta_0)] & \mbox{for model~(a)} \\
                                   ~&~                    \\
(\displaystyle\frac{\eta}{\eta_0})^{2\epsilon_0 \over H}& \mbox{for model~(b)} 
               \end{array}
        \right.
\label{udef}
\ee
and
\be
~~~~~~~T(k) = \left\{ \begin{array}{ll} 
\displaystyle\frac{2k}{\tilde{k}^2}\frac{\Lambda_0}{\epsilon_0}& 
                                                        \mbox{for model~(a)}\\
                                        ~&~                    \\
\displaystyle2\pi R'(0)\frac{\Lambda_0 }{H^2}(-\eta \tilde{k})^{2\epsilon_0/H} 
& \mbox{for model~(b)}. 
               \end{array}
        \right.
\label{tdef}
\ee

%
%
Next we estimate the temporal change of the classicality measures 
introduced in the section 2. 
The classical correlation measure $\delta_{CC}$ and the quantum decoherence 
measure $\delta_{QD}$ are given by 
\bea
\delta_{CC}&=&\sqrt{{AC \over B^2} -1}, \nonumber \\
\delta_{QD}&=&{1 \over 4}(AC-B^2)^{-1/2}.
\eea
Figure 1 is the graph of the numerical estimate of $\delta_{QD}$.
It shows that $\delta_{QD}$ drastically reduces immediately after the 
wave-length exceeds the horizon size.
For the model (a), we observe that $\delta_{QD}$ reaches a plateau stage 
before the wave-length exceeds the horizon size.
Note that
\be
AC-B^2=2 W^2 U^2 \{(D_{10}+D_{11})(D_{30}+D_{31})-{1 \over 4}D_{21}^2\}.
\ee
The dumping factor $U$ reduces the terms $U D_{10}$ and $U D_{30}$ to zero 
after $\epsilon_0 (\eta -\eta_0)$ becomes unity(model (a)) or when 
$(\eta /\eta_0) \ll 1$ is satisfied(model (b)).
Evaluating only the dominant terms, 
we find an asymptotic formula of $\delta_{QD}$ for $|\tilde{k}\eta| \ll 1$:
\be
\delta_{QD}\sim\left\{\begin{array}{ll}
(\epsilon_0/\tilde{k})^{1/2}(\Lambda_0/\tilde{k})^{-1}|\tilde{k} \eta|^{1/2} & \mbox{for model~(a)}\\
                                               ~&~                   \\
const \times \displaystyle\frac{H^2}{\Lambda_0}|\tilde{k}\eta|^{\nu-\epsilon_0/H}
                                                & \mbox{for model~(b)}
                      \end{array}
               \right.  .
\label{qd_asym}
\ee
Here we evaluated 
$R(|\tilde{k}\eta_0|)-R(|\tilde{k}\eta|)-2\tilde\epsilon I_r$ as 
$\tilde{k}/(4\epsilon_0)$ for the model (a).
The measure $\delta_{QD}$ would be one for pure states.  

\par
On the other hand, the change of $\delta_{CC}$ is drawn in Fig.2.
It shows that the initial ground state of the open system is squeezed  
due to the instability induced by the expansion of 
the universe and the classical correlation has been established.
In the asymptotic region $|\tilde{k}\eta |\ll 1$, it decreases as
\be
\delta_{CC}\sim\left\{\begin{array}{ll}
2\pi (\epsilon_0/k_0)
|\tilde{k}\eta|^{5/2} & \mbox{for model~(a)}\\
                                               ~&~                   \\
const \times |\tilde{k}\eta|^{(\nu-\epsilon_0/H)}
                                                & \mbox{for model~(b)}
                      \end{array}
            \right.  .
\label{cc_asym}
\ee
Here we again evaluated 
$R(|\tilde{k}\eta_0|)-R(|\tilde{k}\eta|)-2\tilde\epsilon I_r$ as 
$\tilde{k}/(4\epsilon_0)$ for the model (a).
In this formula, the dissipative coefficient  $\epsilon_0$ appear, but 
the diffusion coefficient  $\Lambda_0$ does not.
From Eqs. \ref{qd_asym} and \ref{cc_asym},  
we conclude that both the classicality conditions are simultaneously satisfied.
This is an example that both the conditions are compatible for unstable 
states, as is discussed using a simple model in Morikawa \cite{morikawa90}.
We can define the squeezing angle $\phi$ as 
\be
{\rm tan}(\phi)\equiv{\beta \over k},
\ee
which means the squeezing direction in the phase space as seen from the Wigner 
representation Eq.\ref{wigner}.
The angle is estimated as 
\be
{\rm tan}(\phi)={1 \over k}{B \over C}
\sim -(\nu+{1 \over 2}){1 \over (-k\eta)}
\ee
for both the models.
The power law in this equation is the same as the pure state except for the amplitude.

%
%
Finally we discuss the power spectrum of the fluctuations.
In the open system, as is seen in the above estimate, 
the criterion of decoherence, $\delta_{QD} \ll 1$, is satisfied due to the 
environment immediately after the wave-length exceeds the horizon 
($|k\eta| \lsim 1$).
The decohered density matrix dominantly contains the statistical fluctuation 
rather than the quantum fluctuations.
Therefore, the open and closed systems are completely different with each 
other with respect to the statistical predictability.
The mean value calculated with  
the density matrix given in Eq.\ref{finalsol} should 
be considered as a statistical average.  
For example, 
full statistically averaged values of $q_k^2$ and its canonical 
conjugate momentum $P_{q_k}^2$ 
are almost the same as those simply derived from the diagonal part of the 
density matrix.
In other words, $<q_k^2>=2C $ and $<P_{q_k}^2>=2A$ 
must be interpreted as the dispersion of values at the different regions 
in the universe.
From Eq.\ref{finalsol}, the power spectrum of the $\phi$ field 
$<\phi_k^2>=2C/a^2$ is explicitly given by 
\be
<\phi_k^2>=U(\eta)<\phi_{\tilde{k}}^2>_0+T(k)<\phi_{\tilde{k}}^2>_0, 
\ee
where $<\phi_{\tilde{k}}^2>_0$ corresponds to the free case in which  
$\Lambda =\epsilon =0$ and shows the Harrison-Zel'dovich spectrum, 
if $\tilde{k}$ is replaced by the wavenumber $k$.
The first term on r.h.s shows that the dissipation effect of the environment 
continues to dump the original quantum fluctuations.
The second term means that the fluctuation effect newly generates 
extra dispersion of $\phi_k$.
The strength of the newly generated dispersion is cooperatively determined by 
the dissipation and fluctuation effects, 
as is found in the expression of $T(k)$, Eq.\ref{tdef}.
The latter effect is enhanced after the wave-length $a/k$ exceeds 
the horizon size and the mode function obtains the non-oscillatory behavior.
As shown in the above estimation, as soon as the wave-length $a/k$ exceeds 
the horizon size, the fluctuation obtains the classical properties.
If the classicalized fluctuation couples with self-gravity, 
it may grow due to the gravitational instability.
In fact, according to the standard theory of the classical fluctuation, 
its amplitude is frozen in the exponential expansion of the universe.
Therefore, we are allowed to identify the seeds of large scale structure as 
the fluctuation just after the wave-length exceeds the horizon
($|k \eta | \lsim 1$).
In general, the power spectrum of the classicalized fluctuation is different 
from that of $<\phi_k^2>_0$ through the k-dependence of the factor $U$ and $T$ 
as well as $\tilde{k}$.
The model (a) is a typical example.
The model (b) is also the example, if $\Lambda_0$ depends on the wavenumber.
On the contrary, 
if we assume that the environment is a thermal bath and the k-mode couples it 
through a bi-linear interaction, we have  
$\Lambda_0 \sim k\epsilon_0$(\cite{caldeira83}).
In this case, T(k) for model (a) has no dependence on the wavenumber.
Therefore, there is also the possibility that 
the environmental effect changes the amplitude of the spectrum 
rather than the power law of k-dependence.

\newsection{Conclusions}
%
%
We calculated the evolution of the density matrix of a scalar field 
in the inflationary universe.
We treated our system as an open system which is affected by dissipation  
and fluctuation effects from the environment.
The density matrix is irreversibly squeezed due to the instability induced 
through the minimal coupling with the space-time dynamics.
Thus, the classical correlation in the phase space develops as 
$\delta_{CC}\to 0$, which   
means that the quantum state makes a transition into the classical state.
In order to discuss the statistical predictability of the state, 
we also examined the evolution of $\delta_{QD}$, which 
measures how a density matrix is mixed.
We revealed that $\delta_{QD}$ 
has drastically decreased at the final stage of inflation.  
Therefore, the density matrix has obtained the capability of statistical 
prediction.
Then we calculated the statistical spectrum of the fluctuation of the 
scalar field.  
We found that the spectrum is different from the power spectrum derived from 
the pure quantum state.  
This is because the environmental effects dump the original 
quantum fluctuation and simultaneously generate additional fluctuation which 
depends on the wave number.

In general, the effect of an environment is non-local with respect to 
time and space.
In our analysis, we did not take into account this non-locality for simplicity.
Further, the master equation for the density matrix (Eq.\ref{fpinf}) 
contains only the second order polynomials of the operators.
It corresponds to the perturbation expansion of the effective action with 
respect to the coupling constant\cite{morikawa86a}.
In order to predict the power spectrum of the  seeds of the universe, we need 
to clarify the k-dependence of the coefficients of dissipation and 
diffusion, 
$\epsilon$ and $\Lambda$, which must be derived from further fundamental 
analysis.
What we have shown using the simple models is that the environmental 
effect cannot be neglected; 
This effect not only makes quantum coherence disappear and 
yields statistical fluctuation, but also inevitably alters the power law 
and the amplitude of the original quantum fluctuations.

Matacz \cite{matacz94} also discussed the power spectrum of the decohered 
scalar field.
He wrote down a squeezed state by  the coherent state representation and 
diagonalized it in non-dynamically {\it ad hoc} way.
His procedure was based on the fact that the coherent state is the most 
robust state to the environmental effects.
For the argument to be correct,  
it must be assumed that 
the relevant observed system is a harmonic oscillator and 
the configuration variable 
and its conjugate momentum exchanges their roles periodically
\cite{unruh89}.
Therefore, the situation he selected implicitly is a reheating stage when 
the inflation expansion has ended and the expansion effect is ignored: 
the potential is not upside-down.
On the other hand, we focused the stage where 
the wave-length steps out the horizon and the coupling with the expansion 
of the universe cannot be neglected.
In the upside-down model which we have advocated, 
the decoherence process must be calculated dynamically.
The asymptotic entropy generation is 
\bea
S&=&-{\rm ln} ({\alpha \over \gamma})-1+O((\alpha/\gamma)) \nonumber\\
 &\sim&\left\{\begin{array}{rl}
             -{1 \over 2}~{\rm ln}|k\eta|  & \mbox{for model(a)}\\
-(\nu-{\epsilon_0 \over H}){\rm ln}|k\eta| & \mbox{for model(b)}
                      \end{array}
       \right. 
\eea
for our upside-down model, while
\be
S\sim-4 {\rm ln}|k\eta| ~~~~~~~~~~~~~~~~~~~~~~~~~~~~~~~~~~
\ee
for the final state in the Matacz model.
The above difference shows that, for the upside-down model, 
the environment does not realize the complete diagonalization 
with respect to the coherent state of the system.

We discussed the classicalization of the quantum field and its power spectrum 
at the stage when the wave-length crosses the horizon size of the 
inflationary universe.
However, it is the classical fluctuation at the first stage of the big 
bang universe that we can know by the observation with the aid of  
the classical self-gravitating theory of the matter. 
Therefore, in oder to predict the statistical properties of the seeds for the 
origin of the structure of the universe, we must further consider 
a) the evolution of the classicalized fluctuation of the super-horizon, 
b) the effect of the reheating stage which occurs at the end of the 
inflationary expansion.
Especially, the reheating is expected to proceed the classicalization of 
the fluctuations.
Therefore, 
we need further research on these problems to clarify the very origin of the 
structure in the universe.

\newpage
\noindent
{\bf Figure Captions}

\noindent
Fig.1
Time evolution of the classicality measure $\delta_{QD}$ is depicted.
It is estimated by the numerical integration.
The environmental effect begins at $\tilde{k}\eta_0=-40$.
Fig.1(a) and (b) correspond to the model (a) and (b), respectively.
The dissipation coefficient, $\epsilon_0/\tilde{k}$ for the model (a) and 
$\epsilon_0/H$ for the model (b) is chosen to be 0.1, 0.05, 0.005.
The diffusion coefficient, $\Lambda_0/\tilde{k}^2$ for the model (a) and 
$\Lambda_0/H^2$ for the model (b), is set to 0.2.

\noindent
Fig.2
Time evolution of the classicality measure $\delta_{CC}$ is depicted.
The parameters are the same as in Fig.1.
Fig.2(a) and (b) correspond to the model (a) and (b), respectively.

\pagebreak

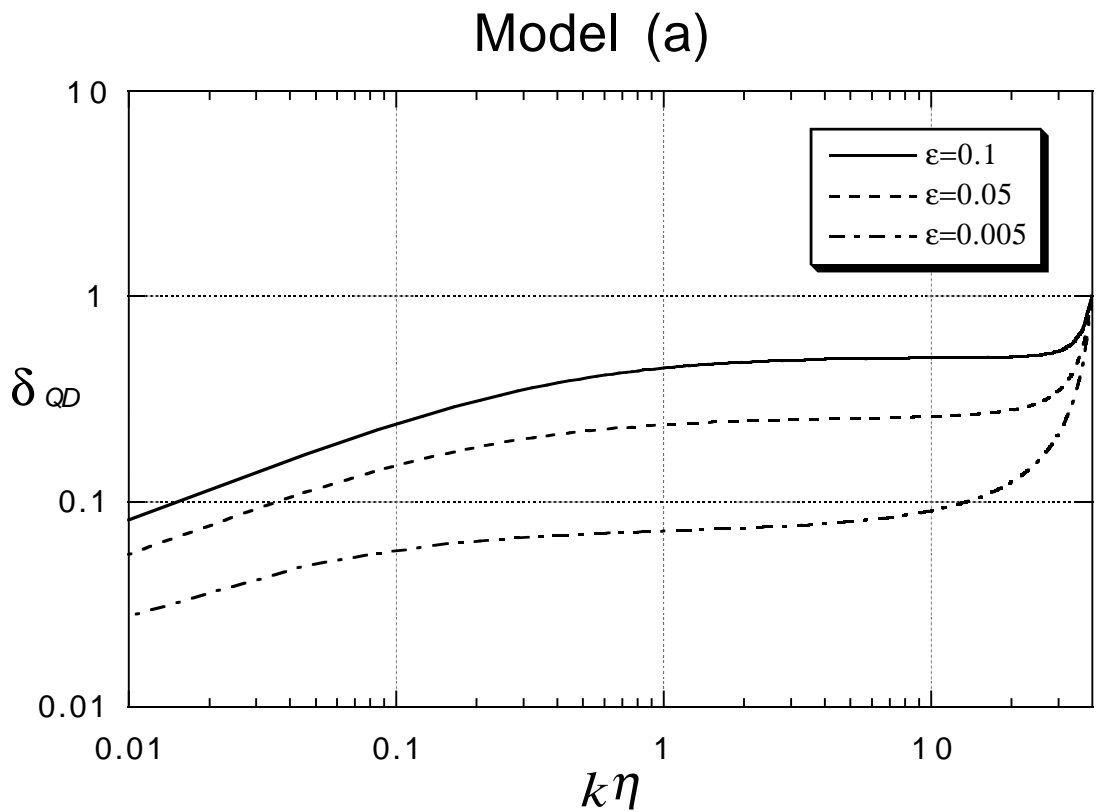
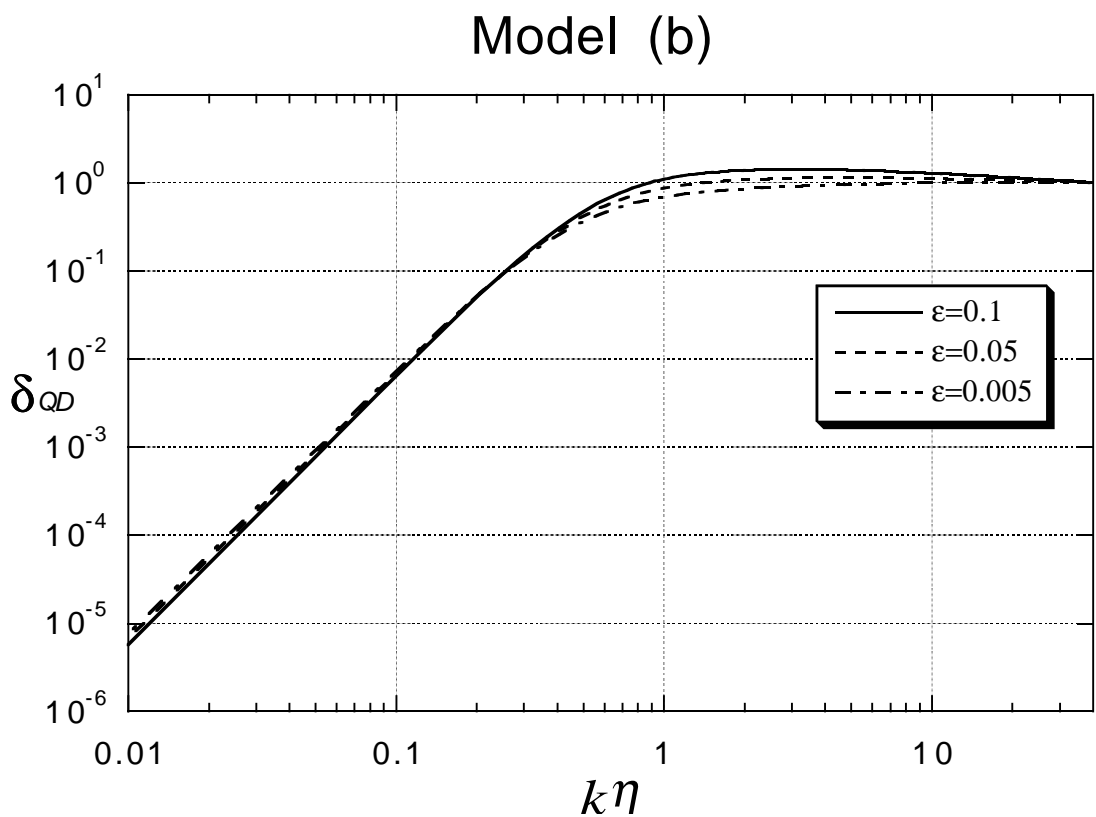
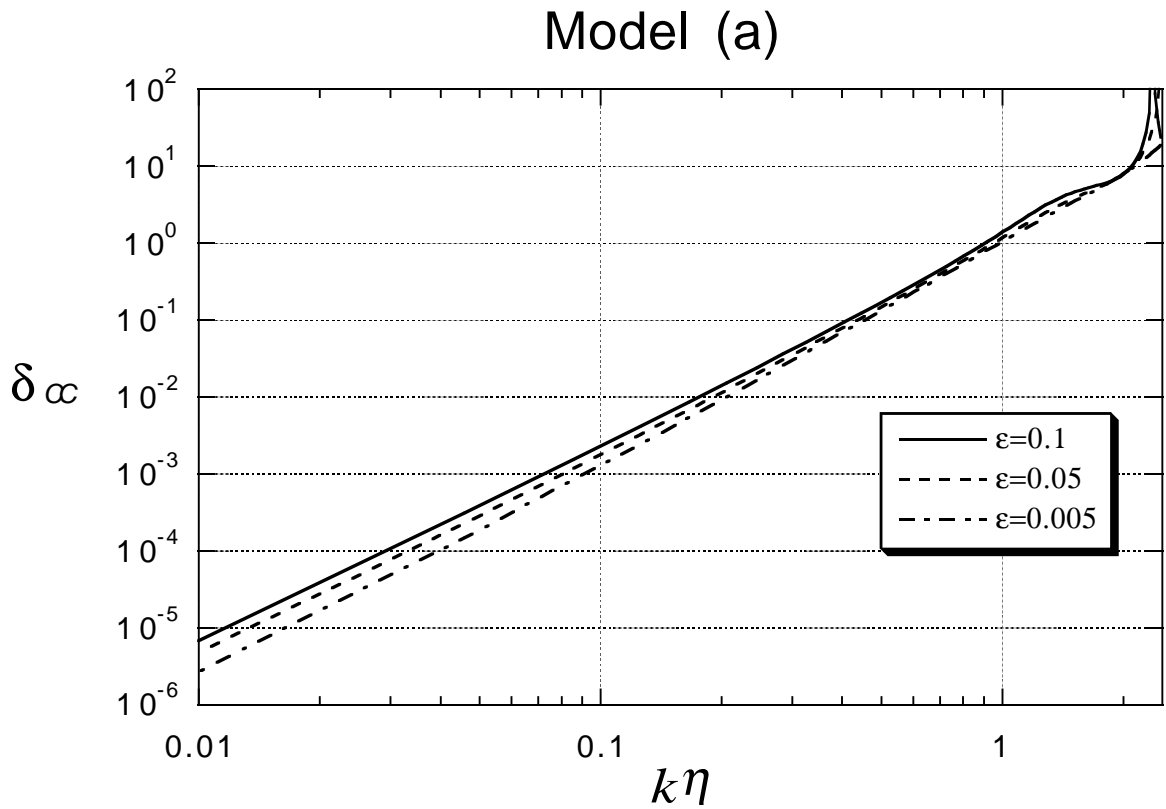
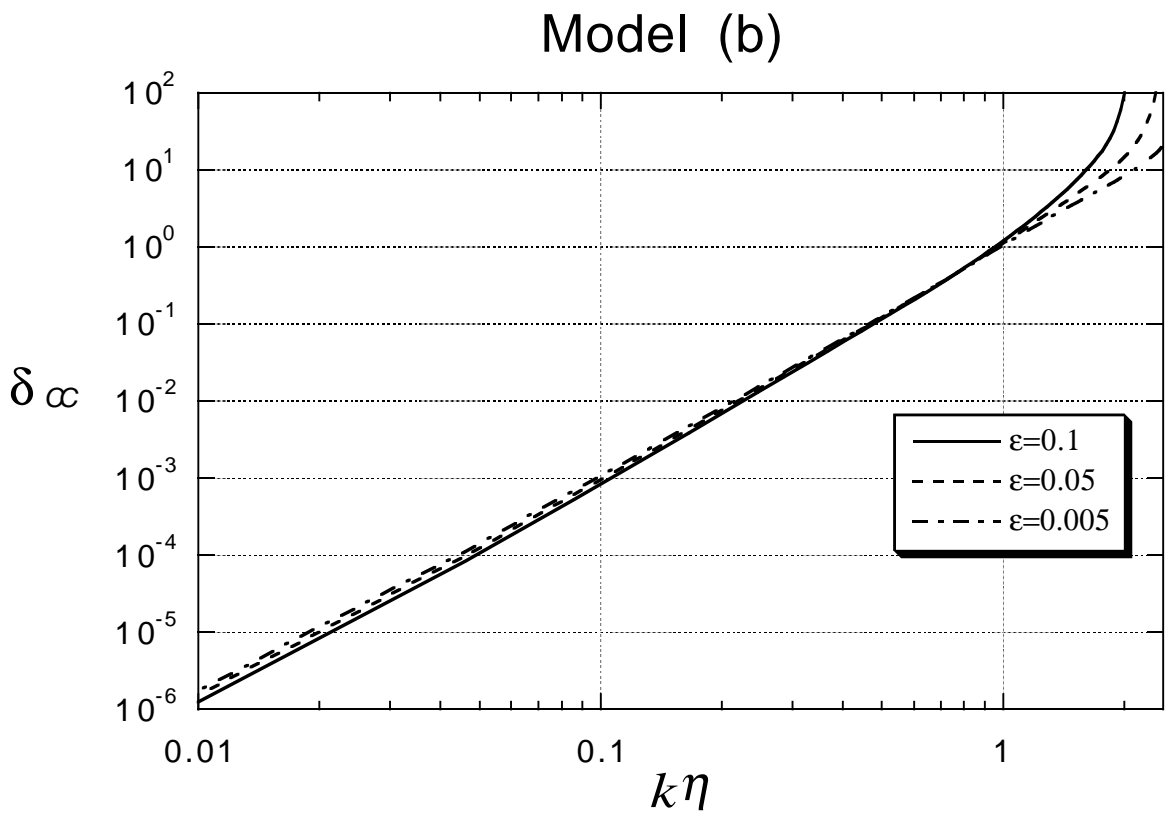
\end{document}